\documentclass[
aps,
pre,
preprint,
longbibliography,
floatfix
]{revtex4-2}

\usepackage[utf8]{inputenc}
\usepackage[T1]{fontenc}
\usepackage{newtxtext, newtxmath}
\usepackage[
colorlinks=true,
urlcolor=purple,
citecolor=purple,
filecolor=purple,
linkcolor=purple
]{hyperref}
\usepackage{amsmath}
\usepackage{graphicx}
\usepackage{xcolor}
\usepackage{tikz-cd}
\usepackage[subfolder]{gnuplottex} 
\usepackage{setspace}
\usepackage{tabularx}

\begin{document}

\title{Smooth and global Ising universal scaling functions}

\author{Jaron Kent-Dobias}
\affiliation{Laboratoire de Physique de l'Ecole Normale Supérieure, Paris, France}

\author{James P.~Sethna}
\affiliation{Laboratory of Atomic and Solid State Physics, Cornell University, Ithaca, NY, USA}

\date\today

\begin{abstract}
  We describe a method for approximating the universal scaling functions for
  the Ising model in a field. By making use of parametric coordinates, the free
  energy scaling function has a polynomial series everywhere. Its form is
  taken to be a sum of the simplest functions that contain the singularities
  which must be present: the Langer essential singularity and the Yang--Lee
  edge singularity. Requiring that the function match series expansions in
  the low- and high-temperature zero-field limits fixes the parametric
  coordinate transformation. For the two-dimensional Ising model, we show that
  this procedure converges exponentially with the order to which the series are
  matched, up to seven digits of accuracy. 
  To facilitate use, we provide Python and Mathematica implementations of the code at both lowest order (three digit) and high accuracy.
\end{abstract}

\maketitle

\section{Introduction}

At continuous phase transitions the thermodynamic properties of physical
systems have singularities. Celebrated renormalization group analyses imply
that not only the principal divergence but entire functions are
\emph{universal}, meaning that they will appear at any critical points that
connect phases of the same symmetries in the same spatial dimension. The study
of these universal functions is therefore doubly fruitful: it provides both a
description of the physical or model system at hand, and \emph{every other
system} whose symmetries, interaction range, and dimension puts it in the same
universality class.

The continuous phase transition in the two-dimensional Ising model is the most
well studied, and its universal thermodynamic functions have likewise received
the most attention. Without a field, an exact solution is known for some
lattice models \cite{Onsager_1944_Crystal}. Precision numeric work both on
lattice models and on the ``Ising'' conformal field theory (related by
universality) have yielded high-order polynomial expansions of those functions,
along with a comprehensive understanding of their analytic properties
\cite{Fonseca_2003_Ising, Mangazeev_2008_Variational, Mangazeev_2010_Scaling}.
In parallel, smooth approximations of the Ising equation of state produce
convenient, evaluable, differentiable empirical functions
\cite{Caselle_2001_The}. Despite being differentiable, these approximations
become increasingly poor when derivatives are taken due to the neglect of
subtle singularities.

This paper attempts to find the best of both worlds: a smooth approximate
universal thermodynamic function that respects the global analytic properties
of the Ising free energy. By constructing approximate functions with the
correct singularities, corrections converge \emph{exponentially} to the true
function.  To make the construction, we review the analytic properties of the
Ising scaling function. Parametric coordinates are introduced that remove
unnecessary singularities that are a remnant of the coordinate choice. The
singularities known to be present in the scaling function are incorporated in
their simplest form. Then, the arbitrary analytic functions that compose those
coordinates are approximated by truncated polynomials whose coefficients are
fixed by matching the series expansions of the universal function.

For the two-dimensional Ising model, this method produces scaling functions
accurate to within $3\times 10^{-4}$ using just the values of the first three
derivatives of the function evaluated at two points, e.g., critical amplitudes
of the magnetization, susceptibility, and first generalized susceptibility.
With six derivatives, it is accurate to about $10^{-7}$. We hope that with some
refinement, this idea might be used to establish accurate scaling functions for
critical behavior in other universality classes, doing for scaling functions
what advances in conformal bootstrap did for critical exponents
\cite{Gliozzi_2014_Critical}. Mathematica and Python implementations will be provided in the supplemental material.

\section{Universal scaling functions}

A renormalization group analysis predicts that certain thermodynamic functions
will be universal in the vicinity of \emph{any} critical point in the Ising
universality class, from perturbed conformal fields to the end of the
liquid--gas coexistence line. Here we will review precisely what is meant by
universal.

Suppose one controls a temperature-like parameter $T$ and a magnetic field-like
parameter $H$, which in the proximity of a critical point at $T=T_c$ and $H=0$
have normalized reduced forms $t=(T-T_c)/T_c$ and $h=H/T$. Thermodynamic
functions are derived from the free energy per site $f=(F-F_c)/N$, which
depends on $t$, $h$, and a litany of irrelevant parameters we will henceforth
neglect.  Explicit renormalization with techniques like the
$\epsilon$-expansion or exact solutions like Onsager's can be used calculated
the flow of these parameters under continuous changes of scale $e^\ell$,
yielding equations of the form
\begin{align} \label{eq:raw.flow}
  \frac{dt}{d\ell}=\frac1\nu t+\cdots
  &&
  \frac{dh}{d\ell}=\frac{\beta\delta}\nu h+\cdots
  &&
  \frac{df}{d\ell}=Df+\cdots
\end{align}
where $D=2$ is the dimension of space and $\nu=1$, $\beta=\frac18$, and
$\delta=15$ are dimensionless constants. The combination
$\Delta=\beta\delta=\frac{15}8$ will appear often. The flow equations are
truncated here, but in general all terms allowed by the symmetries of the
parameters are present on their righthand side. By making a near-identity
transformation to the coordinates and the free energy of the form $u_t(t,
h)=t+\cdots$, $u_h(t, h)=h+\cdots$, and $u_f(f,u_t,u_h)\propto f(t,h)-f_a(t,h)$, one can bring
the flow equations into the agreed upon simplest normal form
\begin{align} \label{eq:flow}
  \frac{du_t}{d\ell}=\frac1\nu u_t
  &&
  \frac{du_h}{d\ell}=\frac{\Delta}\nu u_h
  &&
  \frac{du_f}{d\ell}=Du_f-\frac1{4\pi}u_t^2
\end{align}
which are exact as written \cite{Raju_2019_Normal}. The flow of the
\emph{scaling fields} $u_t$ and $u_h$ is made exactly linear, while that of the
free energy is linearized as nearly as possible. The quadratic term in that
equation is unremovable due to a `resonance' between the value of $\nu$ and the
spatial dimension in two dimensions, while its coefficient is chosen as a
matter of convention, fixing the scale of $u_t$. Here the free energy $f=u_f+f_a$, where $u_f(u_t,u_h)$ is known as the singular part of the free energy, and $f_a(t,h)$ is a non-universal but analytic background free energy.

Solving these equations for $u_f$ yields
\begin{equation}
  \begin{aligned}
    u_f(u_t, u_h)
    &=|u_t|^{D\nu}\mathcal F_\pm(u_h|u_t|^{-\Delta})+\frac{|u_t|^{D\nu}}{8\pi}\log u_t^2 \\
    &=|u_h|^{D\nu/\Delta}\mathcal F_0(u_t|u_h|^{-1/\Delta})+\frac{|u_t|^{D\nu}}{8\pi}\log u_h^{2/\Delta} \\
  \end{aligned}
\end{equation}
where $\mathcal F_\pm$ and $\mathcal F_0$ are undetermined universal scaling functions
related by a change of coordinates \footnote{To connect the results of this
  paper with Mangazeev and Fonseca, one can write $\mathcal
  F_0(\eta)=\tilde\Phi(-\eta)=\Phi(-\eta)+(\eta^2/8\pi) \log \eta^2$ and
$\mathcal F_\pm(\xi)=G_{\mathrm{high}/\mathrm{low}}(\xi)$.}.  The scaling
functions are universal in the sense that any system in the same universality class will share the free energy \eqref{eq:flow}, for suitable analytic functions $u_t$, $u_h$, and analytic background $f_a$ -- the singular behavior is universal up to an analytic coordinate change.
The invariant scaling combinations that appear as the
arguments to the universal scaling functions will come up often, and we will
use $\xi=u_h|u_t|^{-\Delta}$ and $\eta=u_t|u_h|^{-1/\Delta}$.

The analyticity of the free energy at places away from the critical point
implies that the functions $\mathcal F_\pm$ and $\mathcal F_0$ have power-law
expansions of their arguments about zero, the result of so-called Griffiths
analyticity \cite{Griffiths_1967_Thermodynamic}. For instance, when $u_t$ goes to zero for nonzero $u_h$ there is
no phase transition, and the free energy must be an analytic function of its
arguments. It follows that $\mathcal F_0$ is analytic about zero. This is not
the case at infinity: since
\begin{equation}
  \mathcal F_\pm(\xi)
  =\xi^{D\nu/\Delta}\mathcal F_0(\pm \xi^{-1/\Delta})+\frac1{8\pi}\log\xi^{2/\Delta}
\end{equation}
and $\mathcal F_0$ has a power-law expansion about zero, $\mathcal F_\pm$ has a
series like $\xi^{D\nu/\Delta-j/\Delta}$ for $j\in\mathbb N$ at large $\xi$,
along with logarithms. The nonanalyticity of these functions at infinite
argument can be understood as an artifact of the chosen coordinates.

For the scale of $u_t$ and $u_h$, we adopt the same convention as used by
\cite{Fonseca_2003_Ising}. The dependence of the nonlinear scaling variables on
the parameters $t$ and $h$ is system-dependent, and their form can be found for
common model systems (the square- and triangular-lattice Ising models) in the
literature \cite{Mangazeev_2010_Scaling, Clement_2019_Respect}.

\section{Singularities}

\subsection{Essential singularity at the abrupt transition}

In the low temperature phase, the free energy has an essential singularity at
zero field, which becomes a branch cut along the negative-$h$ axis when
analytically continued to negative $h$ \cite{Langer_1967_Theory}. The origin
can be schematically understood to arise from a singularity that exists in the
imaginary free energy of the metastable phase of the model. When the
equilibrium Ising model with positive magnetization is subjected to a small
negative magnetic field, its equilibrium state instantly becomes one with a
negative magnetization. However, under physical dynamics it takes time to
arrive at this state, which happens after a fluctuation containing a
sufficiently large equilibrium `bubble' occurs.

The bulk of such a bubble of radius $R$ lowers the free energy by $2M|H|\pi
R^2$, where  $M$ is the magnetization, but its surface raises the free energy
by $2\pi R\sigma$, where $\sigma$ is the surface tension between the
stable--metastable interface. The bubble is sufficiently large to catalyze the decay of the
metastable state when the differential bulk savings outweigh the surface costs.
This critical bubble occurs with free energy cost
\begin{equation}
  \Delta F_c
    \simeq\frac{\pi\sigma^2}{2M|H|}
    \simeq T\left(\frac{2M_0}{\pi\sigma_0^2}|\xi|\right)^{-1}
\end{equation}
where $\sigma_0=\lim_{t\to0}t^{-\mu}\sigma$ and $M_0=\lim_{t\to0}t^{-\beta}M$
are the critical amplitudes for the surface tension and magnetization at zero
field in the low-temperature phase \cite{Kent-Dobias_2020_Novel}.  In the
context of statistical mechanics, Langer demonstrated that the decay rate is
asymptotically proportional to the imaginary part of the free energy in the
metastable phase, with
\begin{equation}
  \operatorname{Im}F\propto\Gamma\sim e^{-\beta\Delta F_c}\simeq e^{-1/b|\xi|}
\end{equation}
which can be more rigorously derived in the context of quantum field theory
\cite{Voloshin_1985_Decay}. The constant $b=2M_0/\pi\sigma_0^2$ is predicted by
known properties, e.g., for the square lattice $M_0$ and $\sigma_0$ are both
predicted by Onsager's solution \cite{Onsager_1944_Crystal}, but for our
conventions for $u_t$ and $u_h$, $M_0/\sigma_0^2=\bar
s=2^{1/12}e^{-1/8}A^{3/2}$, where $A$ is Glaisher's constant
\cite{Fonseca_2003_Ising}.

\begin{figure}
  \includegraphics{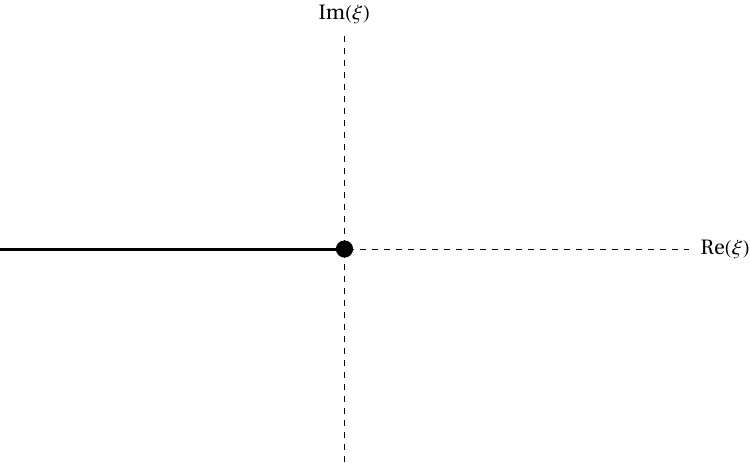}
  \caption{
    Analytic structure of the low-temperature scaling function $\mathcal F_-$
    in the complex $\xi=u_h|u_t|^{-\Delta}\propto H$ plane. The circle
    depicts the essential singularity at the first order transition, while the
    solid line depicts Langer's branch cut.
  } \label{fig:lower.singularities}
\end{figure}

To lowest order, this singularity is a function of the scaling invariant $\xi$
alone. This suggests that it should be considered a part of the
singular free energy, and thus part of the scaling function that composes
it. There is substantial numeric evidence for this as well
\cite{Enting_1980_An, Fonseca_2003_Ising}. We will therefore make the ansatz
that
\begin{equation} \label{eq:essential.singularity}
  \operatorname{Im}\mathcal F_-(\xi+i0)=A_0\Theta(-\xi)\xi e^{-1/b|\xi|}\left[1+O(\xi)\right]
\end{equation}
The linear prefactor can be found through a more careful accounting of the
entropy of long-wavelength fluctuations in the droplet surface
\cite{Gunther_1980_Goldstone, Houghton_1980_The}. In the Ising conformal field
theory, the prefactor is known to be $A_0=\bar s/2\pi$
\cite{Voloshin_1985_Decay, Fonseca_2003_Ising}. The signature of this
singularity in the scaling function is a superexponential divergence in the
series coefficients about $\xi=0$, which asymptotically take the form
\begin{equation} \label{eq:low.asymptotic}
  \mathcal F_-^\infty(m)=\frac{A_0}\pi b^{m-1}\Gamma(m-1)
\end{equation}

\subsection{Yang--Lee edge singularity}

At finite size, the Ising model free energy is an analytic function of
temperature and field because it is the logarithm of a sum of positive analytic
functions. However, it can and does have singularities in the complex plane due
to zeros of the partition function at complex argument, and in particular at
imaginary values of field, $h$. Yang and Lee showed that in the thermodynamic
limit of the high temperature phase of the model, these zeros form a branch cut
along the imaginary $h$ axis that extends to $\pm i\infty$ starting at the
point $\pm ih_{\mathrm{YL}}$ \cite{Yang_1952_Statistical, Lee_1952_Statistical}.
The singularity of the phase transition occurs because these branch cuts
descend and touch the real axis as $T$ approaches $T_c$, with
$h_{\mathrm{YL}}\propto t^{\Delta}$. This implies that the
high-temperature scaling function for the Ising model should have complex
branch cuts beginning at $\pm i\xi_{\mathrm{YL}}$ for a universal constant
$\xi_{\mathrm{YL}}$.

\begin{figure}
  \includegraphics{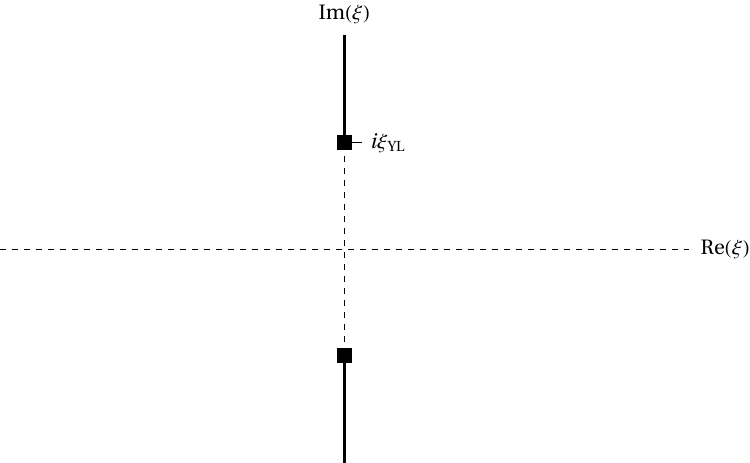}
  \caption{
    Analytic structure of the high-temperature scaling function $\mathcal F_+$
    in the complex $\xi=u_h|u_t|^{-\Delta}\propto H$ plane. The squares
    depict the Yang--Lee edge singularities, while the solid lines depict
    branch cuts.
  } \label{fig:higher.singularities}
\end{figure}

The Yang--Lee singularities, although only accessible with complex fields, are
critical points in their own right, with their own universality class different
from that of the Ising model \cite{Fisher_1978_Yang-Lee}. Asymptotically close
to this point, the scaling function $\mathcal F_+$ takes the form
\begin{equation} \label{eq:yang.lee.sing}
  \mathcal F_+(\xi)
  =A(\xi) +B(\xi)[1+(\xi/\xi_{\mathrm{YL}})^2]^{1+\sigma}+\cdots
\end{equation}
with edge exponent $\sigma=-\frac16$ and $A$ and $B$ analytic functions at
$\xi_\mathrm{YL}$ \cite{Cardy_1985_Conformal, Fonseca_2003_Ising}. This creates
a branch cut stemming from the critical point along the imaginary-$\xi$ axis
with a growing imaginary part
\begin{equation}
  \operatorname{Im}\mathcal F_+(i\xi\pm0)=\pm\tilde A_\mathrm{YL}\Theta(\xi-\xi_\mathrm{YL})(\xi-\xi_\mathrm{YL})^{1+\sigma}[1+O[(\xi-\xi_\mathrm{YL})^2]]
\end{equation}
This results in analytic structure for $\mathcal F_+$ shown in
Fig.~\ref{fig:higher.singularities}. The signature of this in the scaling
function is an asymptotic behavior of the coefficients which goes like
\begin{equation} \label{eq:high.asymptotic}
  \mathcal F_+^\infty(m)=A_\mathrm{YL}2(-1)^{2m}\theta_\mathrm{YL}^{1-\sigma-m}\binom{1-\sigma}{m}
\end{equation}

\section{Parametric coordinates}

The invariant combinations $\xi=u_h|u_t|^{-\Delta}$ or
$\eta=u_t|u_h|^{-1/\Delta}$ are natural variables to describe the scaling
functions, but prove unwieldy when attempting to make smooth approximations.
This is because, when defined in terms of these variables, scaling functions
that have polynomial expansions at small argument have nonpolynomial expansions
at large argument. Rather than deal with the creative challenge of dreaming up
functions with different asymptotic expansions in different limits, we adopt
another coordinate system, in terms of which a scaling function can be defined
that has polynomial expansions in \emph{all} limits.

The Schofield coordinates $R$ and $\theta$ are implicitly defined by
\begin{align} \label{eq:schofield}
  u_t(R, \theta) = R(1-\theta^2)
  &&
  u_h(R, \theta) = R^{\Delta}g(\theta)
\end{align}
where $g$ is an odd function whose first zero lies at $\theta_0>1$
\cite{Schofield_1969_Parametric}. We take
\begin{align} \label{eq:schofield.funcs}
  g(\theta)=\left(1-\frac{\theta^2}{\theta_0^2}\right)\sum_{i=0}^\infty g_i\theta^{2i+1}.
\end{align}
This means that $\theta=0$ corresponds to the high-temperature zero-field line,
$\theta=1$ to the critical isotherm at nonzero field, and $\theta=\theta_0$ to
the low-temperature zero-field (phase coexistence) line.
In practice the infinite series in \eqref{eq:schofield.funcs} cannot be
entirely fixed, and it will be truncated at finite order.

\begin{figure}
  \include{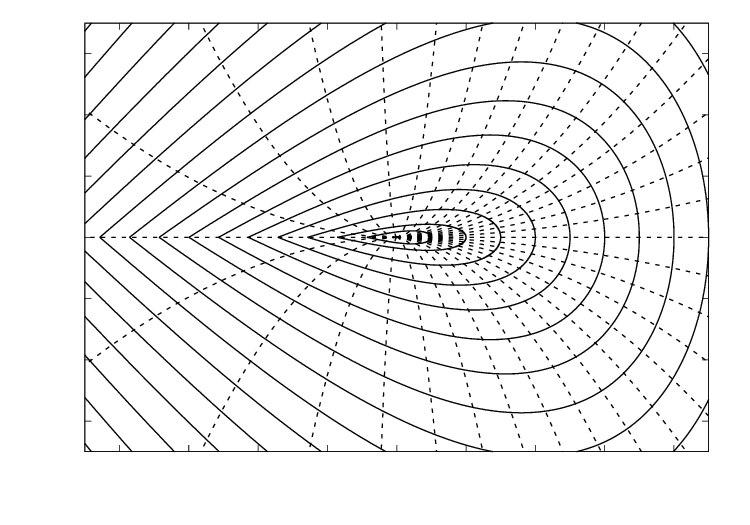}
  \caption{
    Example of the parametric coordinates. Solid lines are of constant
    $R=\frac12,1,\ldots,8\frac12$  and dashed lines are of constant
    $\theta=\pm0\theta_0,\frac1{16}\theta_0,\ldots,\theta_0$ for $g(\theta)$ taken
    from the $n=6$ entry of Table \ref{tab:fits}.
  } \label{fig:schofield}
\end{figure}

One can now see the convenience of these coordinates. Both invariant scaling
combinations depend only on $\theta$, as
\begin{align}
  \xi=u_h|u_t|^{-\Delta}=\frac{g(\theta)}{|1-\theta^2|^{\Delta}} &&
  \eta=u_t|u_h|^{-1/\Delta}=\frac{1-\theta^2}{|g(\theta)|^{1/\Delta}}
\end{align}
Moreover, both scaling variables have polynomial expansions in $\theta$ near
zero, with
\begin{align}
  &\xi= g'(0)\theta+\cdots  && \text{for $\theta\simeq0$}\\
  &\xi=g'(\theta_0)(\theta_0^2-1)^{-\Delta}(\theta-\theta_0)+\cdots && \text{for $\theta\simeq\theta_0$}
  \\
  &\eta=-2(\theta-1)g(1)^{-1/\Delta}+\cdots && \text{for $\theta\simeq1$}.
\end{align}
Since the scaling functions $\mathcal F_\pm(\xi)$ and $\mathcal F_0(\eta)$ have
polynomial expansions about small $\xi$ and $\eta$, respectively, this implies
both will have polynomial expansions in $\theta$ everywhere.

\begin{figure}
  \includegraphics{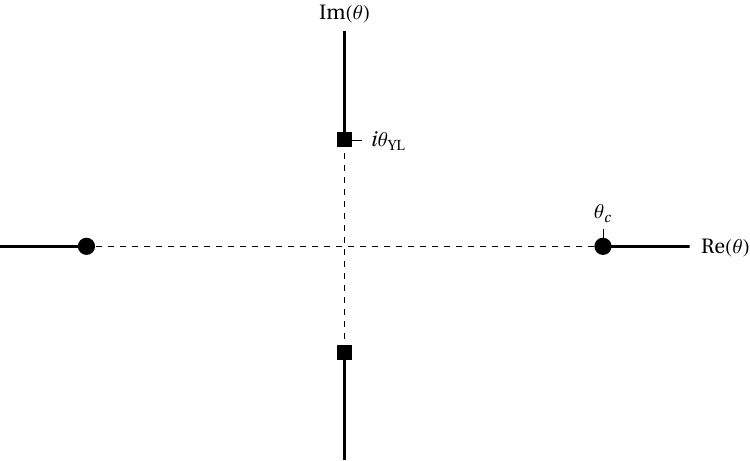}
  \caption{
    Analytic structure of the global scaling function $\mathcal F$ in the
    complex $\theta$ plane. The circles depict essential singularities of the
    first order transitions, the squares the Yang--Lee singularities, and the
    solid lines depict branch cuts.
  } \label{fig:schofield.singularities}
\end{figure}

Therefore, in Schofield coordinates one expects to be able to define a global
scaling function $\mathcal F(\theta)$ which has a polynomial expansion in its
argument for all real $\theta$ by
\begin{equation}
  u_f(R,\theta)=R^{D\nu}\mathcal F(\theta)+(1-\theta^2)^2\frac{R^2}{8\pi}\log R^2
\end{equation}
For small $\theta$, $\mathcal F(\theta)$ will
resemble $\mathcal F_+$, for $\theta$ near one it will resemble $\mathcal F_0$,
and for $\theta$ near $\theta_0$ it will resemble $\mathcal F_-$. This can be
seen explicitly using the definitions \eqref{eq:schofield} to relate the above
form to the original scaling functions, giving
\begin{equation} \label{eq:scaling.function.equivalences.2d}
  \begin{aligned}
    \mathcal F(\theta)
    &=|1-\theta^2|^{D\nu}\mathcal F_\pm\left[g(\theta)|1-\theta^2|^{-\Delta}\right]
    +\frac{(1-\theta^2)^2}{8\pi}\log(1-\theta^2)^2\\
    &=|g(\theta)|^{D\nu/\Delta}\mathcal F_0\left[(1-\theta^2)|g(\theta)|^{-1/\Delta}\right]
    +\frac{(1-\theta^2)^2}{8\pi}\log g(\theta)^{2/\Delta}
  \end{aligned}
\end{equation}
This leads us
to expect that the singularities present in these functions will likewise be
present in $\mathcal F(\theta)$. The analytic structure of this function is
shown in Fig.~\ref{fig:schofield.singularities}. Two copies of the Langer
branch cut stretch out from $\pm\theta_0$, where the equilibrium phase ends,
and the Yang--Lee edge singularities are present on the imaginary-$\theta$
line (because  $\mathcal F$ has the same symmetry in $\theta$
as $\mathcal F_+$ has in $\xi$).

The location of the Yang--Lee edge singularities can be calculated directly
from the coordinate transformation \eqref{eq:schofield}. Since $g(\theta)$ is
an odd real polynomial for real $\theta$, it is imaginary for imaginary
$\theta$. Therefore,
\begin{equation} \label{eq:yang-lee.theta}
  i\xi_{\mathrm{YL}}=\frac{g(i\theta_{\mathrm{YL}})}{(1+\theta_{\mathrm{YL}}^2)^{-\Delta}}
\end{equation}
The location $\theta_0$ is not fixed by any principle.

\section{Functional form for the parametric free energy}

As we have seen in the previous sections, the unavoidable singularities in the
scaling functions are readily expressed as singular functions in the imaginary
part of the free energy.

Our strategy follows. First, we take the singular imaginary parts of the
scaling functions $\mathcal F_{\pm}(\xi)$ and truncate them to the lowest order
accessible under polynomial coordinate changes of $\xi$. Then, we constrain
the imaginary part of $\mathcal F(\theta)$ to have this simplest form,
implicitly defining the analytic parametric coordinate change $g(\theta)$. Third, we perform a
Kramers--Kronig type transformation to establish an explicit form for the real
part of $\mathcal F(\theta)$ involving a second analytic function $G(\theta)$. Finally, we make good on the constraint made in
the second step by fitting the coefficients of $g(\theta)$ and $G(\theta)$ to reproduce the
correct known series coefficients of $\mathcal F_{\pm}$.

This success of this stems from the commutative diagram below. So long as the
application of Schofield coordinates and the Kramers--Kronig relation can be
said to commute, we may assume we have found correct coordinates for the
simplest form of the imaginary part to be fixed later by the real part.
\[
  \begin{tikzcd}[row sep=large, column sep = 9em]
  \operatorname{Im}\mathcal F_\pm(\xi) \arrow{r}{\text{Kramers--Kronig in $\xi$}} \arrow[]{d}{\text{Schofield}} & \operatorname{Re}\mathcal F_{\pm}(\xi) \arrow{d}{\text{Schofield}} \\%
  \operatorname{Im}\mathcal F(\theta) \arrow{r}{\text{Kramers--Kronig in $\theta$}}& \operatorname{Re}\mathcal F(\theta)
\end{tikzcd}
\]
We require that, for $\theta\in\mathbb R$
\begin{equation} \label{eq:imaginary.abrupt}
  \operatorname{Im}\mathcal F(\theta+0i)=\operatorname{Im}\mathcal F_0(\theta+0i)=C_0[\Theta(\theta-\theta_0)\mathcal I(\theta)-\Theta(-\theta-\theta_0)\mathcal I(-\theta)]
\end{equation}
where
\begin{equation}
  \mathcal I(\theta)=(\theta-\theta_0)e^{-1/B(\theta-\theta_0)}
\end{equation}
reproduces the essential singularity in \eqref{eq:essential.singularity}.
Independently, we require for $\theta\in\mathbb R$
\begin{equation}
  \operatorname{Im}\mathcal F(i\theta+0)
  =\operatorname{Im}\mathcal F_\mathrm{YL}(i\theta+0)
  =\frac12C_\mathrm{YL}\left[
    \Theta(\theta-\theta_\mathrm{YL})(\theta-\theta_\mathrm{YL})^{1+\sigma}
    -\Theta(\theta+\theta_\mathrm{YL})(\theta+\theta_\mathrm{YL})^{1+\sigma}
  \right]
\end{equation}
Fixing these requirements for the imaginary part of $\mathcal F(\theta)$ fixes
its real part up to an analytic even function $G(\theta)$, real for real $\theta$.

\begin{figure}
  \includegraphics{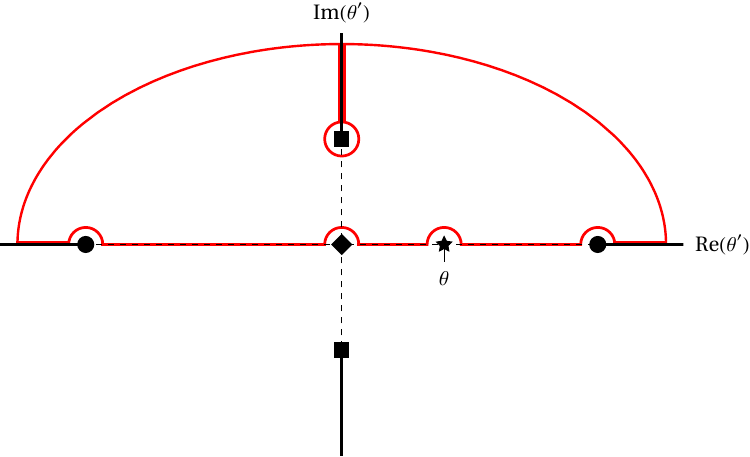}
  \caption{
    Integration contour over the global scaling function $\mathcal F$ in the
    complex $\theta$ plane used to produce the dispersion relation. The
    circular arc is taken to infinity, while the circles around the
    singularities are taken to zero.
  } \label{fig:contour}
\end{figure}

To find the real part of the nonanalytic part of the scaling function, we make
use of the identity
\begin{equation}
  0=\oint_{\mathcal C}d\vartheta\,\frac{\mathcal F(\vartheta)}{\vartheta^2(\vartheta-\theta)}
\end{equation}
where $\mathcal C$ is the contour in Figure \ref{fig:contour}. The integral is
zero because there are no singularities enclosed by the contour. The only
nonvanishing contributions from this contour as the radius of the semicircle is
taken to infinity are along the real line and along the branch cut in the upper
half plane. For the latter contributions, the real parts of the integration up
and down cancel out, while the imaginary part doubles. This gives
\begin{equation}
  \begin{aligned}
    0&=\left[\int_{-\infty}^\infty+\lim_{\epsilon\to0}\left(\int_{i\infty-\epsilon}^{i\theta_{\mathrm{YL}}-\epsilon}+\int^{i\infty+\epsilon}_{i\theta_{\mathrm{YL}}+\epsilon}\right)\right]
      d\vartheta\,\frac{\mathcal F(\vartheta)}{\vartheta^2(\vartheta-\theta)} \\
     &=\int_{-\infty}^\infty d\vartheta\,\frac{\mathcal F(\vartheta)}{\vartheta^2(\vartheta-\theta)}
     +2i\int_{i\theta_{\mathrm{YL}}}^{i\infty}d\theta'\,\frac{\operatorname{Im}\mathcal F(\vartheta)}{\vartheta^2(\vartheta-\theta)} \\
     &=-i\pi\frac{\mathcal F(\theta)}{\theta^2}+\mathcal P\int_{-\infty}^\infty d\vartheta\,\frac{\mathcal F(\vartheta)}{\vartheta^2(\vartheta-\theta)}
     +2i\int_{i\theta_{\mathrm{YL}}}^{i\infty}d\vartheta\,\frac{\operatorname{Im}\mathcal F(\vartheta)}{\vartheta^2(\vartheta-\theta)}
  \end{aligned}
\end{equation}
where $\mathcal P$ is the principle value.  In principle one would need to
account for the residue of the pole at zero, but since its order is less than
two and $\mathcal F(0)=\mathcal F'(0)=0$, this evaluates to zero. Rearranging, this gives
\begin{equation}
  \mathcal F(\theta)
  =\frac{\theta^2}{i\pi}\mathcal P\int_{-\infty}^\infty d\vartheta\,\frac{\mathcal F(\vartheta)}{\vartheta^2(\vartheta-\theta)}
  +\frac{2\theta^2}\pi\int_{i\theta_{\mathrm{YL}}}^{i\infty}d\vartheta\,\frac{\operatorname{Im}\mathcal F(\theta')}{\vartheta^2(\vartheta-\theta)}
\end{equation}
Taking the real part of both sides, we find
\begin{equation}
  \operatorname{Re}\mathcal F(\theta)
  =\frac{\theta^2}{\pi}\mathcal P\int_{-\infty}^\infty d\vartheta\,\frac{\operatorname{Im}\mathcal F(\vartheta)}{\vartheta^2(\vartheta-\theta)}
  -\frac{2\theta^2}\pi\int_{\theta_{\mathrm{YL}}}^{\infty}d\vartheta\,\frac{\operatorname{Im}\mathcal F(i\vartheta)}{\vartheta(\vartheta^2+\theta^2)}
\end{equation}
Because the real part of $\mathcal F$ is even, the imaginary part must be odd. Therefore
\begin{equation} \label{eq:dispersion}
  \operatorname{Re}\mathcal F(\theta)
  =\frac{\theta^2}{\pi}
  \int_{\theta_0}^\infty d\vartheta\,\frac{\operatorname{Im}\mathcal F(\vartheta)}{\vartheta^2}\left(\frac1{\vartheta-\theta}+\frac1{\vartheta+\theta}\right)
  -\frac{2\theta^2}\pi\int_{\theta_{\mathrm{YL}}}^{\infty}d\vartheta\,\frac{\operatorname{Im}\mathcal F(i\vartheta)}{\vartheta(\vartheta^2+\theta^2)}
\end{equation}
Evaluating these ordinary integrals, we find for $\theta\in\mathbb R$
\begin{equation}
  \operatorname{Re}\mathcal F(\theta)=\operatorname{Re}\mathcal F_0(\theta)+\mathcal F_\mathrm{YL}(\theta)+G(\theta)
\end{equation}
where
\begin{equation} \label{eq:2d.real.Fc}
  \operatorname{Re}\mathcal F_0(\theta)
  =C_0[\mathcal R(\theta)+\mathcal R(-\theta)]
\end{equation}
where $\mathcal R$ is given by the function
\begin{equation}
  \mathcal R(\theta)
  =\frac1\pi\left[
    \theta_0e^{1/B\theta_0}\operatorname{Ei}(-1/B\theta_0)
    +(\theta-\theta_0)e^{-1/B(\theta-\theta_0)}\operatorname{Ei}(1/B(\theta-\theta_0))
  \right]
\end{equation}
and
\begin{equation}
  \mathcal F_{\mathrm{YL}}(\theta)=2C_\mathrm{YL}\left[2(\theta^2+\theta_\mathrm{YL}^2)^{(1+\sigma)/2}\cos\left((1+\sigma)\tan^{-1}\frac\theta{\theta_\mathrm{YL}}\right)-\theta_\mathrm{YL}^{1+\sigma}\right]
\end{equation}
We have also included the analytic part $G$, which we assume has a simple
series expansion
\begin{equation} \label{eq:analytic.free.enery}
  G(\theta)=\sum_{i=1}^\infty G_i\theta^{2i}
\end{equation}
From the form of the real part, we can infer the form of $\mathcal F$ that is
analytic for the whole complex plane except at the singularities and branch
cuts previously discussed.
For $\theta\in\mathbb C$, we take
\begin{equation}
  \mathcal F(\theta)=\mathcal F_0(\theta)+\mathcal F_{\mathrm{YL}}(\theta)+G(\theta),
\end{equation}
where
\begin{equation}
  \mathcal F_0(\theta)=C_0\left\{
    \mathcal R(\theta)+\mathcal R(-\theta)
    +i\operatorname{sgn}(\operatorname{Im}\theta)[\mathcal I(\theta)-\mathcal I(-\theta)]
  \right\}
\end{equation}

\section{Fitting}

The scaling function has a number of free parameters: the position $\theta_0$
of the abrupt transition, prefactors in front of singular functions from the
abrupt transition and the Yang--Lee point, the coefficients in the analytic
part $G$ of the scaling function, and the coefficients in the undetermined
coordinate function $g$.

The other parameters $B$, $C_0$, $\theta_{YL}$, and $C_{YL}$ are determined or
further constrained by known properties. For $\theta>\theta_0$, the form
\eqref{eq:essential.singularity} can be expanded around $\theta=\theta_0$ to
yield
\begin{equation}
  \begin{aligned}
    \operatorname{Im}u_f
    &\simeq A_0 u_t(\theta)^{D\nu}\xi(\theta)\exp\left\{\frac1{b\xi(\theta)}\right\} \\
    &=A_0R^{D\nu}(\theta_0^2-1)^{D\nu}\xi'(\theta_0)(\theta-\theta_0)
    \exp\left\{\frac1{b\xi'(\theta_0)}\left(\frac1{\theta-\theta_0}
      -\frac{\xi''(\theta_0)}{2\xi'(\theta_0)}\right)
      \right\}\left(1+O[(\theta-\theta_0)^2]\right)
  \end{aligned}
\end{equation}
Comparing this with the requirement \eqref{eq:imaginary.abrupt}, we find that
\begin{equation}
  B=-b\xi'(\theta_0)=-b\frac{g'(\theta_0)}{(\theta_0^2-1)^{1/\Delta}}
\end{equation}
and
\begin{equation}
  \begin{aligned}
    C_0&=A_0t(\theta_0^2-1)^{D\nu}\xi'(\theta_0)\exp\left\{
    -\frac{\xi''(\theta_0)}{2b\xi'(\theta_0)^2}
  \right\} \\
       &=
       A_0(\theta_0^2-1)^{D\nu-\Delta}g'(\theta_0)
       \exp\left\{-\frac1b\left(\frac{(\theta_0^2-1)^\Delta g''(\theta_0)}{2g'(\theta_0)^2}-\frac{2\Delta(\theta_0^2-1)^{\Delta - 1}\theta_0}{g'(\theta_0)}
       \right)\right\}
  \end{aligned}
\end{equation}
fixing $B$ and $C_0$. Similarly, \eqref{eq:yang-lee.theta} puts a constraint on
the value of $\theta_\mathrm{YL}$, while the known amplitude of the Yang--Lee
branch cut fixes the value of $C_\mathrm{YL}$ by
\begin{equation}
  \begin{aligned}
    u_f
    &\simeq A_\mathrm{YL}|u_h(\theta)|^{D\nu/\Delta}(\eta_{\mathrm YL}-\eta(\theta))^{1+\sigma} \\
    &=A_\mathrm{YL}R^{D\nu}|g(i\theta_\mathrm{YL})|^{D\nu/\Delta}[-\eta'(i\theta_\mathrm{YL})]^{1+\sigma}(\theta-i\theta_\mathrm{YL})^{1+\sigma}\left(1+O[(\theta-i\theta_\mathrm{YL})^2]\right)\\
    &\simeq R^{D\nu}\mathcal F_\mathrm{YL}(\theta)
    =C_\mathrm{YL}R^{D\nu}(\theta-i\theta_\mathrm{YL})^{1+\sigma}\left(1+O[(\theta-i\theta_\mathrm{YL})^2]\right)
\end{aligned}
\end{equation}
\begin{equation}
  C_\mathrm{YL}=A_\mathrm{YL}|g(i\theta_\mathrm{YL})|^{D\nu/\Delta}\left[\frac{-\eta'(i\theta_\mathrm{YL})}{2i\theta_\mathrm{YL}}\right]^{1+\sigma}
\end{equation}
where $A_\mathrm{YL}=-1.37(2)$ and $\xi_\mathrm{YL}=0.18930(5)$
\cite{Fonseca_2003_Ising}. Because these parameters are not known exactly,
these constraints are added to the weighted sum of squares rather than
substituted in.

This leaves as unknown variables the positions $\theta_0$ and
$\theta_{\mathrm{YL}}$ of the abrupt transition and Yang--Lee edge singularity,
the amplitude $C_\mathrm{YL}$ of the latter, and the unknown functions $G$ and
$g$. We determine these approximately by iteration in the polynomial order at
which the free energy and its derivative matches known results, shown in
Table~\ref{tab:data}. We write as a cost function the difference between the
known series coefficients of the scaling functions $\mathcal F_\pm$ and the
series coefficients of our parametric form evaluated at the same points,
$\theta=0$ and $\theta=\theta_0$, weighted by the uncertainty in the value of
the known coefficients or by a machine-precision cutoff, whichever is larger.
We also add the difference between the predictions for $A_\mathrm{YL}$ and
$\xi_\mathrm{YL}$ and their known numeric values, again weighted by their
uncertainty. In order to encourage convergence, we also add weak residuals
$j!g_j$ and $j!G_j$ encouraging the coefficients of the analytic functions $g$
and $G$ in \eqref{eq:schofield.funcs} and \eqref{eq:analytic.free.enery} to
stay small.  This can be interpreted as a prior which expects these functions
to be analytic, and therefore have series coefficients which decay with a
factorial.

A Levenberg--Marquardt algorithm is performed on the cost function to find a parameter combination which minimizes it. As larger polynomial order in the
series are fit, the truncations of $G$ and $g$ are extended to higher order so
that the codimension of the fit is constant.
We performed this procedure starting at $n=2$ (matching the scaling
function at the low and high temperature zero field points to quadratic order), up
through $n=6$. At higher order we began to have difficulty minimizing the cost.
The resulting fit coefficients can be found in Table \ref{tab:fits}.

\begin{figure}
  \include{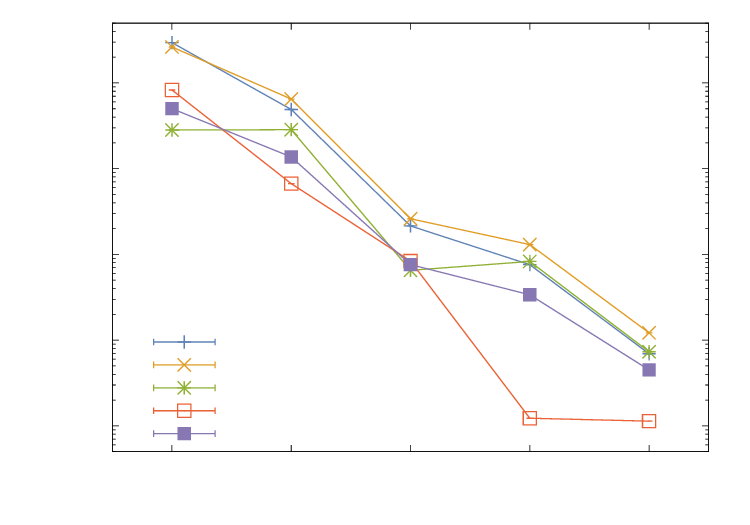}
  \caption{
    The error in the $m$th derivative of the scaling function $\mathcal F_0$
    with respect to $\eta$ evaluated at $\eta=0$, as a function of the
    polynomial order $n$ at which the scaling function was fit. The point
    $\eta=0$ corresponds to the critical isotherm at $T=T_c$ and $H>0$, roughly
    midway between the two limits used in the fit, at $H=0$ and $T$ above and
    below $T_c$. Convergence here should reflect overall convergence of our
    scaling function at all $\theta$.} \label{fig:error}
\end{figure}

\begin{figure}
  \include{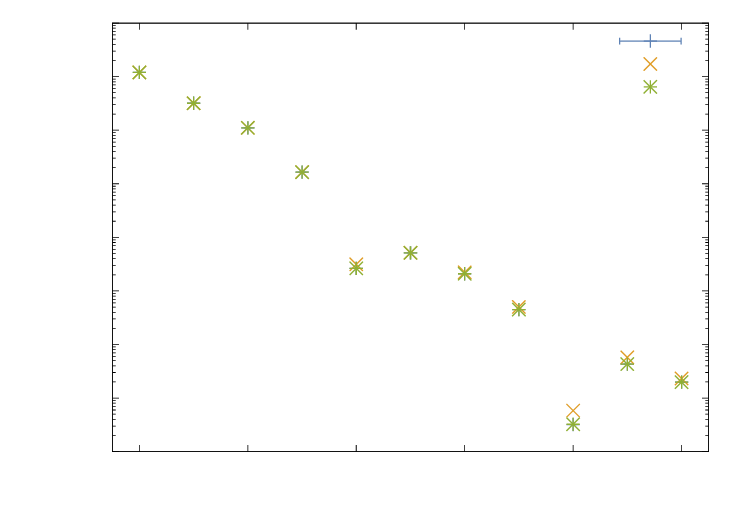}
  \caption{
    The series coefficients for the scaling function $\mathcal F_0$ as a
    function of polynomial order $m$. The numeric values are from Table
    \ref{tab:data} and are partially obscured by the other data.
  } \label{fig:phi.series}
\end{figure}

Precise results exist for the value of the scaling function and its derivatives
at the critical isotherm, or equivalently for the series coefficients of the
scaling function $\mathcal F_0$. Since we do not use these coefficients in our
fits, the error in the approximate scaling functions and their derivatives can
be evaluated by comparison to their known values at the critical isotherm, or
$\theta=1$.  The difference between the numeric values of the coefficients
$\mathcal F_0^{(m)}$ and those predicted by the iteratively fit scaling
functions are shown in Fig.~\ref{fig:error}. For the values for which we were
able to make a fit, the error in the function and its first several derivatives
appear to trend exponentially towards zero in the polynomial order $n$. The
predictions of our fits at the critical isotherm can be compared with the
numeric values to higher order in Fig.~\ref{fig:phi.series}, where the absolute
values of both are plotted.
\begin{table}
  \begin{tabular}{r|lll}
    \multicolumn1{c|}{$m$} &
      \multicolumn{1}{c}{$\mathcal F_-^{(m)}$} &
      \multicolumn{1}{c}{$\mathcal F_0^{(m)}$} &
      \multicolumn1c{$\mathcal F_+^{(m)}$} \\
      \hline
    0 &
      \hphantom{$-$}0 &
      $-1.197\,733\,383\,797\ldots$ &
      \hphantom{$-$}0 \\
    1 &
      $-1.357\,838\,341\,707\ldots$ &
      \hphantom{$-$}$0.318\,810\,124\,891\ldots$ &
      \hphantom{$-$}0 \\
    2 &
      $-0.048\,953\,289\,720\ldots$ &
      \hphantom{$-$}$0.110\,886\,196\,683(2)$ &
      $-1.845\,228\,078\,233\ldots$ \\
    3 &
      \hphantom{$-$}$0.038\,863\,932(3)$ &
      $-0.016\,426\,894\,65(2)$ &
      \hphantom{$-$}0 \\
    4 &
      $-0.068\,362\,119(2)$ &
      $-2.639\,978(1)\times10^{-4}$ &
      \hphantom{$-$}$8.333\,711\,750(5)$ \\
    5 &
      \hphantom{$-$}$0.183\,883\,70(1)$ &
      \hphantom{$-$}$5.140\,526(1)\times10^{-4}$ &
      \hphantom{$-$}0 \\
    6 &
      $-0.659\,171\,4(1)$ &
      \hphantom{$-$}$2.088\,65(1)\times 10^{-4}$ &
      $-95.168\,96(1)$ \\
    7 &
      \hphantom{$-$}$2.937\,665(3)$ &
      \hphantom{$-$}$4.481\,9(1)\times10^{-5}$ &
      \hphantom{$-$}0 \\
    8 &
      $-15.61(1)$ &
      \hphantom{$-$}$3.16\times10^{-7}$ &
      \hphantom{$-$}1457.62(3) \\
    9 &
      \hphantom{$-$}96.76 &
      $-4.31\times10^{-6}$ &
      \hphantom{$-$}0 \\
    10 &
      $-679$ &
      $-1.99\times10^{-6}$ &
      $-25\,891(2)$ \\
    11 &
    \hphantom{$-$}$5.34\times10^3$ & &
      \hphantom{$-$}0 \\
    12 &
      $-4.66\times10^4$ & &
      \hphantom{$-$}$5.02\times10^5$ \\
    13 &
      \hphantom{$-$}$4.46\times10^5$ & &
      \hphantom{$-$}0 \\
    14 &
      $-4.66\times10^6$ & &
      $-1.04\times10^7$
  \end{tabular}
  \caption{
    Known series coefficients for the universal scaling functions. Those with
    trailing dots are known exactly or have closed integral representations.
    Those with listed uncertainties are taken from Mangazeev \textit{et al.}\
    \cite{Mangazeev_2008_Variational}. Those without are taken from Fonseca
    \textit{et al.}, and are assumed to be accurate to within their last digit
    \cite{Fonseca_2003_Ising}.
  } \label{tab:data}
\end{table}

\begin{table}
  \singlespacing
  \raggedright
  \begin{tabular}{|c|lllllll}
    \hline
    \multicolumn{1}{|c|}{$n$} &
      \multicolumn{1}{c}{$\theta_0$} &
      \multicolumn{1}{c}{$\theta_\mathrm{YL}$} &
      \multicolumn{1}{c}{$C_\mathrm{YL}$} &
      \multicolumn{1}{c}{$G_1$} &
      \multicolumn{1}{c}{$G_2$} &
      \multicolumn{1}{c}{$G_3$} &
      \multicolumn{1}{c}{$G_4$} \\
    \hline
    2 &
      1.14841 &
      0.989667 &
      $-0.172824$ &
      $-0.310183$ &
      0.247454 \\
    3 &
      1.25421 &
      0.602056 &
      $-0.385664$ &
      $-0.352751$ &
      0.258243 \\
    4 &
      1.31649 &
      0.640019 &
      $-0.356397$ &
      $-0.355055$ &
      0.234659 &
      $-0.00190837$ \\
    5 &
      1.34032 &
      0.623811 &
      $-0.380029$ &
      $-0.351275$ &
      0.237046 &
      $-0.00731973$ \\
    6 &
      1.36261 &
      0.646215 &
      $-0.355764$ &
      $-0.352058$ &
      0.233166 &
      $-0.00664903$ &
      $-0.00168991$ \\
      \hline
  \end{tabular}
  \begin{tabular}{|c|llllll}
    \hline
    $n$ &
      \multicolumn{1}{c}{$g_0$} &
      \multicolumn{1}{c}{$g_1$} &
      \multicolumn{1}{c}{$g_2$} &
      \multicolumn{1}{c}{$g_3$} &
      \multicolumn{1}{c}{$g_4$} &
      \multicolumn{1}{c}{$g_5$} \\
    \hline
    2 &
      0.373691 &
      $-0.0216363$ \\
    3 &
      0.448379 &
      $-0.0220323$ &
      \hphantom{$-$}0.000222006 \\
    4 &
      0.441074 &
      $-0.0348177$ &
      \hphantom{$-$}0.000678173 &
      $-0.0000430514$ \\
    5 &
      0.443719 &
      $-0.0460994$ &
      $-0.000745834$ &
      \hphantom{$-$}0.0000596688 &
      $-0.00000440308$ \\
    6 &
      0.438453 &
      $-0.0531270$ &
      $-0.00391478$ &
      $-0.000408016$ &
      \hphantom{$-$}0.0000262629 &
      $-0.00000109745$ \\
    \hline
  \end{tabular}
  \caption{
    Free parameters in the fit of the parametric coordinate transformation and
    scaling form to known values of the scaling function series coefficients
    for $\mathcal F_\pm$. The fit at stage $n$ matches those coefficients up to
    and including order $n$. Uncertainty estimates are difficult to quantify directly.
  } \label{tab:fits}
\end{table}

\begin{figure}
  \include{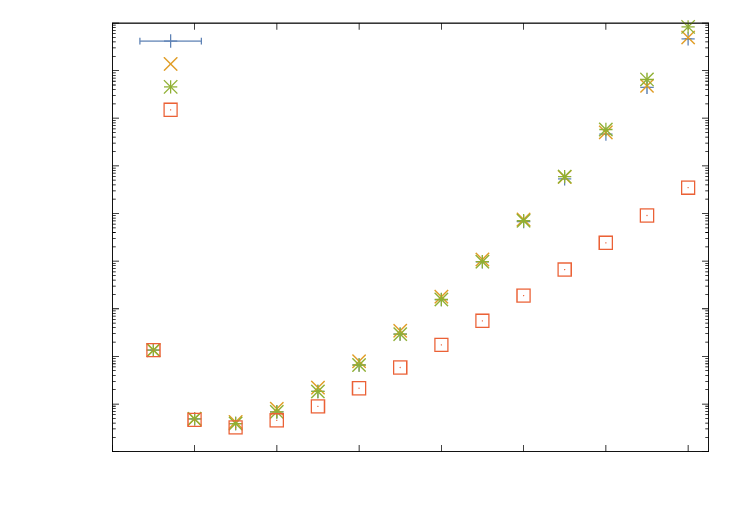}
  \caption{
    The series coefficients for the scaling function $\mathcal F_-$ as a
    function of polynomial order $m$. The numeric values are from Table
    \ref{tab:data}, and those of Caselle \textit{et al.} are from the most
    accurate scaling function listed in \cite{Caselle_2001_The}. The deviation at high polynomial order illustrates the lack of the essential singularity in the form of Caselle \textit{et al.}.
  } \label{fig:glow.series}
\end{figure}

\begin{figure}
  \include{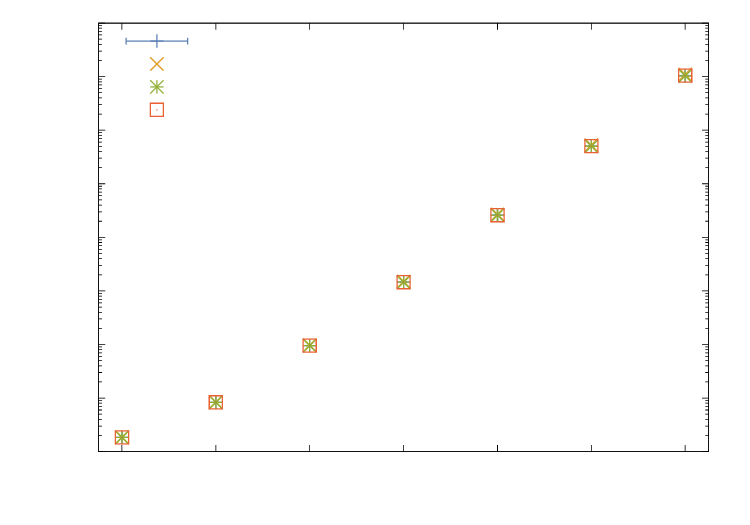}
  \caption{
    The series coefficients for the scaling function $\mathcal F_+$ as a
    function of polynomial order $m$. The numeric values are from Table
    \ref{tab:data}, and those of Caselle \textit{et al.} are from the most
    accurate scaling function listed in \cite{Caselle_2001_The}. Note all agree
    well for $H$ near zero, $T > T_c$, as does the function of Caselle
    \textit{et al}.
  } \label{fig:ghigh.series}
\end{figure}

Even at $n=2$, where only seven unknown parameters have been fit, the results
are accurate to within $3\times10^{-4}$. This approximation for the scaling
functions also captures the singularities at the high- and low-temperature
zero-field points well. A direct comparison between the magnitudes of the
series coefficients known numerically and those given by the approximate
functions is shown for $\mathcal F_-$ in Fig.~\ref{fig:glow.series}, for
$\mathcal F_+$ in Fig.~\ref{fig:ghigh.series}, and for $\mathcal F_0$ in
Fig.~\ref{fig:phi.series}.

\begin{figure}
  \include{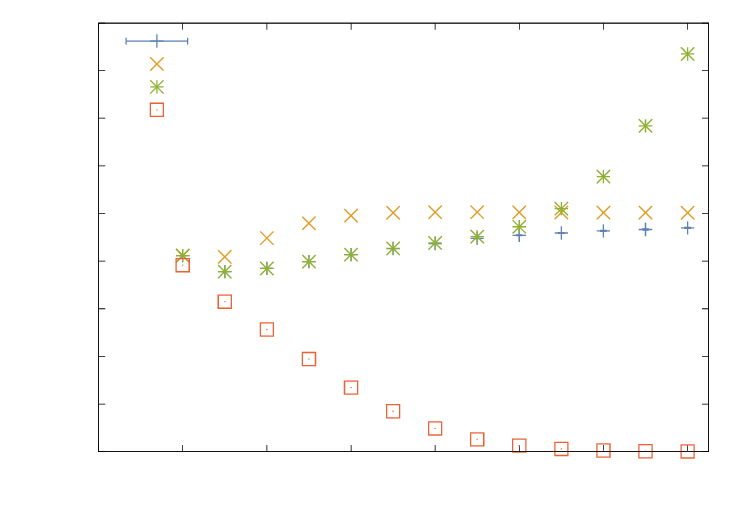}
  \caption{
    The series coefficients for the scaling function $\mathcal F_-$ as a
    function of polynomial order $m$, rescaled by their asymptotic limit
    $\mathcal F_-^\infty(m)$ from \eqref{eq:low.asymptotic}. The numeric values
    are from Table \ref{tab:data}, and those of Caselle \textit{et al.} are
    from the most accurate scaling function listed in \cite{Caselle_2001_The}. Note that our $n=6$ fit generates significant deviations in polynomial coefficients $m$ above around 10. 
  } \label{fig:glow.series.scaled}
\end{figure}

\begin{figure}
  \include{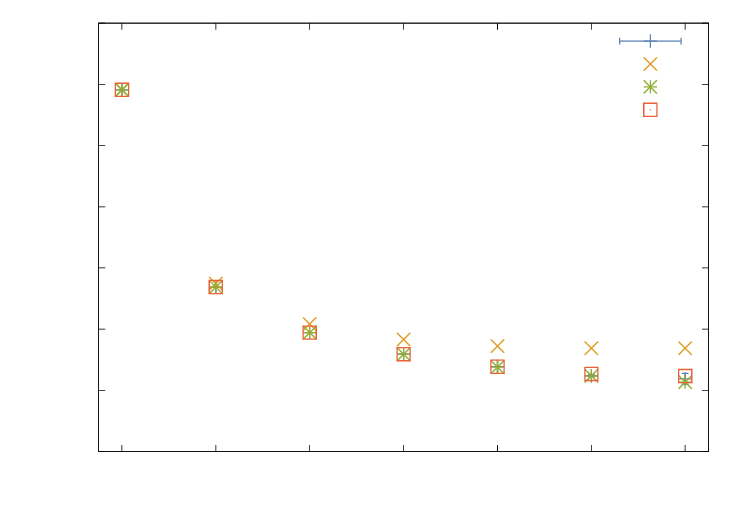}
  \caption{
    The series coefficients for the scaling function $\mathcal F_+$ as a
    function of polynomial order $m$, rescaled by their asymptotic limit
    $\mathcal F_+^\infty(m)$ from \eqref{eq:high.asymptotic}. The numeric
    values are from Table \ref{tab:data}, and those of Caselle \textit{et al.}
    are from the most accurate scaling function listed in \cite{Caselle_2001_The}.
  } \label{fig:ghigh.series.scaled}
\end{figure}

Also shown are the ratio between the series in $\mathcal F_-$ and $\mathcal
F_+$ and their asymptotic behavior, in Fig.~\ref{fig:glow.series.scaled} and
Fig.~\ref{fig:ghigh.series.scaled}, respectively. While our functions have the
correct asymptotic behavior by construction, for $\mathcal F_-$ they appear to
do poorly in an intermediate regime which begins at larger order as the order
of the fit becomes larger. This is due to the analytic part of the scaling
function and the analytic coordinate change, which despite having small
high-order coefficients as functions of $\theta$ produce large intermediate
derivatives as functions of $\xi$. We suspect that the nature of the truncation
of these functions is responsible, and are investigating modifications that
would converge better.
Notice that this infelicity does not appear to cause significant errors in the function $\mathcal F_-(\theta)$ or its low order derivatives, as evidenced by the convergence in Fig.~\ref{fig:error}.

\begin{figure}
  \include{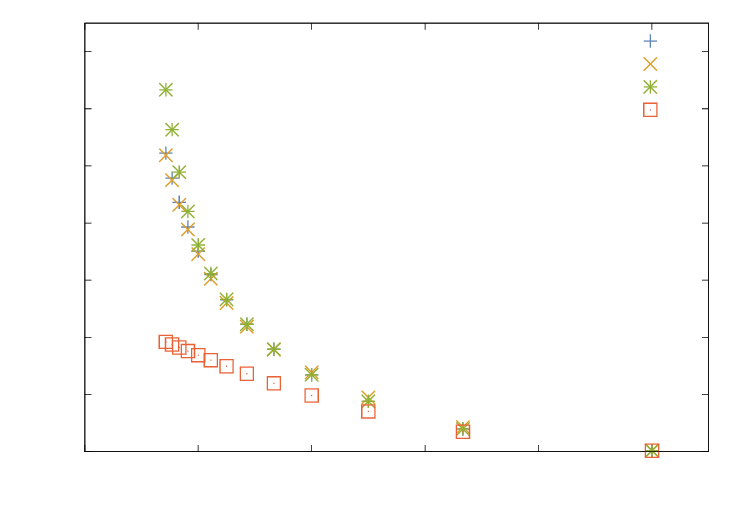}
  \caption{
    Sequential ratios of the series coefficients of the scaling function
    $\mathcal F_-$ as a function of inverse polynomial order $m$. The
    extrapolated $y$-intercept of this plot gives the radius of convergence of
    the series, which should be zero due to the essential singularity (as seen
    in the known numeric values and in this work). Caselle \textit{et al.} do
    not incorporate the essential singularity.
  } \label{fig:glow.radius}
\end{figure}

Besides reproducing the high derivatives in the series, the approximate
functions defined here feature the appropriate singularity at the abrupt
transition. Fig.~\ref{fig:glow.radius} shows the ratio of subsequent series
coefficients for $\mathcal F_-$ as a function of the inverse order, which
should converge in the limit of $m\to0$ to the inverse radius of convergence
for the series. Approximations for the function without the explicit
singularity have a nonzero radius of convergence, where both the numeric data
and the approximate functions defined here show the appropriate divergence in
the ratio.

\section{Outlook}

We have introduced explicit approximate functions forms for the two-dimensional
Ising universal scaling function in the relevant variables. These functions are
smooth to all orders, include the correct singularities, and appear to converge
exponentially to the function as they are fixed to larger polynomial order.

This method, although spectacularly successful, could be improved. It becomes difficult to fit the
unknown functions at progressively higher order due to the complexity of the
chain-rule derivatives, and we find an inflation of predicted coefficients in our higher-precision fits. These problems may be related to the precise form and
method of truncation for the unknown functions.

The successful smooth description of the Ising free energy produced in part by
analytically continuing the singular imaginary part of the metastable free
energy inspires an extension of this work: a smooth function that captures the
universal scaling \emph{through the coexistence line and into the metastable
phase}. The functions here are not appropriate for this except for a small
distance into the metastable phase, at which point the coordinate
transformation becomes untrustworthy. In order to do this, the parametric
coordinates used here would need to be modified so as to have an appropriate
limit as $\theta\to\infty$.

\begin{acknowledgments}
  The authors would like to thank Tom Lubensky, Andrea Liu, and Randy Kamien
  for helpful conversations. The authors would also like to think Jacques Perk
  for pointing us to several insightful studies. JPS thanks Jim Langer for past
  inspiration, guidance, and encouragement. This work was supported by NSF
  grants DMR-1312160 and DMR-1719490. JK-D is supported by the Simons
  Foundation Grant No.~454943.
\end{acknowledgments}

\bibliography{ising_scaling}

\end{document}